\documentclass{ws-procs975x65}

\usepackage{latexsym}
\usepackage{amssymb}
\usepackage{amsmath}
\usepackage{float}
\usepackage{indentfirst}
\usepackage{array}
\usepackage{xspace}

\usepackage[unicode=true]{hyperref}

\usepackage[mathcal]{euscript}

\newcommand{\ba}{\begin{array}}
\newcommand{\ea}{\end{array}}
\newcommand{\be}{\begin{equation}}
\newcommand{\ee}{\end{equation}}
\newcommand{\bea}{\begin{eqnarray}}
\newcommand{\eea}{\end{eqnarray}}
\newcommand{\e}{\mathrm{e}}

\newcommand{\lb}{\label}

\begin{document}
%\begin{flushright}DTP-MSU/12-13
%\end{flushright}

\title{ HORIZON REGULARITY AND DILATON COUPLING\\ QUANTIZATION IN EMD DYONS}
\author{D.\,V.\,GAL'TSOV }
\address{\parbox{10cm}{\noindent\rule{0cm}{0.4cm}{} Faculty
of Physics, Moscow State University, 119899, Moscow, Russia.}  \\
\email{galtsov@phys.msu.ru}}
%\author{A. N. AUTHOR}
%\address{Group, Laboratory, Street, ...}
\begin{abstract}
Four-dimensional Einstein-Maxwell-Dilaton theory admits
asymptotically flat extremal dyonic solutions for an infinite
discrete sequence of the coupling constant values. The quantization
condition arises as consequence of regularity of the dilaton
function at the event horizon. These dyons satisfy the no-force
condition and have flat reduced three spaces like true BPS
configurations, but no supersymmetric embeddings are known except
for some cases of lower values of the coupling sequence.
\end{abstract}

%\bodymatter
\section{Introduction}
Charged  black holes in four-dimensional Einstein-Maxwell-dilaton
(EMD) gravity  exhibit different features depending on the value of
the dilaton coupling constant $a$ entering the Maxwell term
$\e^{-2a\phi}F^2$ in the Lagrangian. Several particular values of
$a$ have higher-dimensional supergravity origin. For $a=0$ the
dilaton decouples and the theory reduces to the Einstein-Maxwell
(EM) system, which is the bosonic part of $N=2,\,D=4$ supergravity.
In this case the extremal dyons, defined geometrically as black
holes with the degenerate event horizon, saturate the supergravity
Bogomol'nyi-Prasad-Sommerfield (BPS) bound, and have $AdS_2\times
S^2$ horizon with zero Hawking temperature. For $a\neq 0$ only the
static purely electric or magnetic black are known analytically
\cite{Gibbons:1982ih,GiWi,Gibbons:1987ps,GHS}. These solutions
generically have two horizons,  the inner one being singular, so in
the extremality limit the event horizon becomes a null singularity
with vanishing Beckenstein-Hawking entropy and finite temperature
(small black holes). For $a=(p/(p+2))^{1/2}$ small black holes may
be interpreted as compactified regular non-dilatonic $p$-branes in
$(4+p)$-dimensional EM theory \cite{Gibbons:1994vm}. The value $a=1$
corresponds to $N=4,\,D=4$ supergravity or dimensionally reduced
heterotic string effective action, in this case the dyonic solutions
are also known \cite{Gibbons:1982ih,Gibbons:1987ps} which are
non-singular in the extremal limit and  possess $AdS_2\times S^2$
horizons. The last particular case, $a=\sqrt{3}$, corresponds to
dimensionally reduced $N=2,\,D=5$ supergravity; in this case the
static dyon solutions  also have the $AdS_2\times S^2$ horizon
structure in the extremality limit.

The rotating dyonic solutions are known analytically only for  $a=0$
and $a=\sqrt{3}$ \cite{Clement:1986bt,Rasheed:1995zv}. In the first
case it is the Kerr-Newman solution of the EM theory, while in the
second  these were derived  using the three-dimensional sigma-model
on the symmetric space $SL(3,R)/SO(2,1)$, corresponding to vacuum
five-dimensional gravity. EMD theories with these two values of the
dilaton coupling exhaust the set of models reducing to
three-dimensional sigma-models on coset spaces
\cite{Galtsov:1995mb}, so from this reasoning there are no
indications on any particular  status of EMD theories with other
$a$. Meanwhile, as was shown numerically by Poletti, Twamley and
Wiltshire \cite{Poletti:1995yq}, the values $a=0,\,\sqrt{3}$ are
just the two lowest members $n=1,\, 2$ of the ``triangular''
sequence of dilaton couplings
 \be \lb{an}
a_n=\sqrt{n(n+1)/2}\,,
  \ee
selecting EMD theories in which numerical non-extremal dyonic
solutions exist with two horizons and admit the extremal limit.

Some attempts are known to explore possibility of supersymmetric
embedding of the EMD theory with arbitrary $a$. Using Witten-Nester
construction, Gibbons et al. \cite{Gibbons:1993xt}, were able to
derive the BPS-like inequality for arbitrary $a$. Meanwhile, as was
later shown by Nozawa and Shiromizu \cite{Nozawa:2010rf}, the
corresponding Killing spinor equations  do not imply the initial
bosonic equations as the integrability condition, unless
$a=0,\,\sqrt{3}$. So the question whether there is some
supersymmetry underlying the rule (\ref{an}) still remains open.

\section{The setup}
 We choose the Einstein-Maxwell-dilaton lagrangian in the form
 \be L=
R  - 2(\partial \phi)^2   - \e^{- 2a\phi}
  F^2\,,
 \label{action}
 \ee
and assume the static ansatz for the metric and the Maxwell one-form
:
 \begin{align}\label{ansatzM}
& ds^2 = -\e^{2 \delta}\, N\,  dt^2 +
\frac{dr^2}{N} +  R^2\,  \left(d\theta^2  + \sin^2 \theta d\phi^2\right)\;,& \\
& A =-f\,dt-P\,\cos \theta\ d\phi\;,&
\end{align}
where $P $ is the magnetic charge.  The functions $f$, $\phi$, $N$,
$R$ and $\delta$ depend on the radial variable $r$ only. The
equations of motions can be readily found from the reduced
one-dimensional lagrangian
 \bea \label{L} &&\mathcal{L} = 2k \e^{\delta} + 2\e^{-\delta} \left (\e^{2\delta} N R\right)'
R' -2\ \phi'^2 \e^{\delta} N R^{2}   + 2\ \e^{-2 a\phi} \left (f'^2
R^{2} \e^{-\delta} - \frac{P^2 \e^{\delta}}{R^{2}}\right)\;. \eea
Two particular gauges are relevant: one is $\delta=0$ in which some
exact solutions are known, another is $R=r$, which is suitable for
analytic derivation of the quantization rules and for numerical
calculations. Imposing the coordinate condition and solving the
Maxwell equations in the gauge $R=r$, we find:
$$ f' =
 \frac{Q\;\e^{\delta+2 a\phi}}{r^{2}}\,,$$ where $Q$ is the integration
 constant (the electric charge). The remaining dilaton and Einstein's
equations then read:
\begin{align}
 & \left(N \varphi' \e^{\delta} r^{2} \right)' = \frac{2a
 \e^\delta |PQ|}{ r^{2}}\sinh(2a\varphi)  \;,&\label{eqP}\\
 & \delta' =\varphi'^2 r\;,& \label{eqN}\\
 &\e^{ -\delta}\  \left (\!\e^{\delta} \!N r \right)'=
  1   -
 \frac{2|PQ| \cosh(2a\varphi)}{ r^{2}} \,,& \label{eqD}
\end{align}
where the  shifted dilaton function is introduced
$$ \varphi=\phi-(\ln z)/2a\,, \qquad z=\left|\frac{P}{Q}\right|.$$
The third Einstein equation is related to (\ref{eqN}-\ref{eqD})  via
the Bianchi identity. This system of equations possess a discrete
electric-magnetic duality \be \lb{sdu} P\leftrightarrow Q\,,\qquad
\varphi\leftrightarrow -\varphi \,.\ee

Asymptotic flatness (AF) implies $N(\infty)=1\,,\,
\delta({\infty})=0$ with the next to leading terms
 \begin{align} & N  \sim  1 - \frac{2M}{r}+
\left(2|QP| \cosh(2a\varphi) + \Sigma^2\right)\frac{1}{r^{2}} , \label{AFN}\\
 &\e^\delta \sim 1-\frac{\Sigma^2}{2 r^{2}} ,\label{AFD}\\
 &\phi\sim \phi_\infty + \frac{\Sigma}{r} , \label{AFP}
 \end{align}
where $ M,\,\Sigma$ and $\phi_\infty$ are free parameters of the
local series solution. As expected, for global solutions the dilaton
charge $\Sigma$ is not an independent quantity: integrating the Eq.
(\ref{eqP}) from $r_h$ to $\infty$, one obtains the sum rule:
\be\label{sigmaint}
 \Sigma =2a|QP|\int_{r_h}^{\infty}
 \frac{\e^\delta\sinh{2a\varphi}}{r^{2}}dr\;.
 \ee

It can be shown that for the AF solutions with the degenerate
horizon   there is a second constraint on the charges, namely  the
no-force condition. First, from the equations of motion one can
deduce that the quadratic form
 \be
I= \left( \frac12 N^2 \e^{2\delta} \varphi'^2  + \frac{1}{8}
\e^{-2\delta}(N\ \e^{2\delta})'^2 \right)r^{4}-
 |QP|N \e^{2\delta}
\cosh(2a\varphi)
 \ee
is conserved on shell:
$
\frac{dI}{dr}=0\,,
$
(similar expression in the gauge $\delta =0$ was given in
 \cite{Poletti:1995yq}).
Substituting here $r=r_h$, which solves two equations $N=0=N'$ in
the case of the degenerate horizon, one finds that this integral has
zero value $I=0$. Then substituting the asymptotic expansions
(\ref{AFN}-\ref{AFP}) we obtain from $I=0$ the no-force condition:
 \be
 M^2+\Sigma =Q_\infty^2+P_\infty^2\,, \lb{no-force}
 \ee
where $Q_\infty=Q\e^{2a\phi_\infty}, P_\infty=P\e^{-2a\phi_\infty}.$
\section{Exact solutions}
An exact extremal dyon solution  with $a=1$ in this gauge reads:
\begin{align}
&\e^{-2\delta}=1+\frac{\Sigma^2}{r^2}\,,\quad
 \e^{2\varphi}=\left|\frac{Q}{P}\right|\cdot
 \frac{\sqrt{r^2+\Sigma^2}+\Sigma}{\sqrt{r^2+\Sigma^2}-\Sigma}\,,\nonumber\\
 & N=
\left(1-\frac{2M}{r^2}\sqrt{r^2+\Sigma^2}+\frac{Q^2+P^2+\Sigma^2}{r^2}\right)\,,
\end{align} It has an event horizon at $r_h=\sqrt{M^2-\Sigma^2}$ and the
dilaton charge satisfying $\lb{Si1} 2M\Sigma=P^2-Q^2\,.
 $
  This solution saturates the  fake BPS bound
\cite{Gibbons:1993xt}:
 \be M \geq \frac{1}{\sqrt{1+a^2}}\sqrt{Q^2 +
P^2}\,. \label{BPS}
 \ee

Another exact dyon solution is known
\cite{Gibbons:1982ih,Gibbons:1987ps} for $n=2,\, (a=\sqrt{3})$ . The
corresponding dilaton charge satisfies an interpolating formula
 \be
 \frac{Q^2}{\Sigma-a M}+\frac{P^2}{\Sigma+a
 M}=\frac{1+a^2}{2a^2}\Sigma\,, \label{consan}
 \ee
 together with (\ref{Si1}) (this is valid only for $a=a_n \,,
 n=1,\,2$).

Further information about higher $n$ dyons can be extracted from
known exact solutions for singly charges black holes. This provides
us with some knowledge about the limiting point $z=0$, an opposite
limit $z=\infty$ can be explored via the discrete electric-magnetic
duality (\ref{sdu})). These solutions are simpler in the gauge
$\delta=0$, the corresponding equations of motion being
 \bea
&&\left(N \varphi' R^{2} \right)' = \frac{2a
 |PQ|}{ R^{2}}\sinh(2a\varphi)  \;, \\
&& R'' +\varphi'^2 R=0\;,\\
&&  \left (N R \right)'=
  k - \Lambda R^{2}  -
 \frac{2|PQ| \cosh(2a\varphi)}{ R^{2}} \,.
\eea The electrically  charged solution  valid for all $a$ in the
general non-extremal case reads:
 \begin{align}\lb{sta}
R^2=\rho^2f_-^{1-\gamma}\,,\quad N=f_+f_-^{\gamma}\,,\quad
\e^{2a\varphi}=f_-^{\frac{2a}{1+a^2}}\,,\quad
\gamma=\frac{1-a^2}{1+a^2}
 \,,
 \end{align}
with $ f_\pm=1-r_\pm/\rho$, where we denoted the radial coordinate
as $\rho$ to distinguish it from $r$ in the gauge used in the Eqs.
(\ref{eqP}-\ref{eqD}). The mass, the electric charge and the dilaton
charge are related to $r_\pm$ via
 \be\lb{char0}
r_+ r_-=\frac{2Q^2}{1+\gamma},\,\quad 2M=r_++\gamma r_-\,,\quad
\Sigma = -\frac{a}{\Delta} r_-\,.
 \ee
Note that our system of equations has special points of two kinds:
the zeroes of the function $R(\rho)$, which correspond to the
curvature singularity, and the zeroes of $N(\rho)$, for which $R\neq
0$; these are regular points. Generic behavior of the metric
functions and the dilaton near the curvature singularity is
non-analytic, while in the second case it {\em is} analytic. This is
clearly seen in the solution (\ref{sta}) which has two zeroes of
$N(\rho)$ at $\rho=r_{\pm}$, with $r_-$ being a zero of $R(\rho)$
too, i.e., singular. The dilaton is analytic at the regular horizon
 $\rho=r_+$, but non-analytic in the singularity
$\rho=r_-$. This can be expected for more general dyon solutions as
well.

In the extremal limit $r_+=r_-$ the dilaton is therefore singular at
the horizon, but this does not influence its asymptotic behavior. So
the dilaton charge  is still finite and given by (\ref{char0}). In
the extremal limit it reads:
 \be \lb{SiQ}\Sigma = -\frac{a\,Q}{\sqrt{\Delta}} \,.
 \ee This expression is expected to match the
corresponding limit of dyonic solutions.

\section{Coupling quantization}
 Now let us look for extremal dyon solutions with the regular
horizon for general dilaton coupling constant. It turns out that the
{\em local} power series solution valid in the vicinity of the
degenerate horizon implies a constraint on the dilaton coupling
constant. Coming back to the gauge $R=r$, in which the curvature
singularity is at $r=0$, and assuming that the horizon radius $r_h$
satisfying $N(r_h)=N'(r_h)=0$ is finite, we will have in the leading
order:
 \bea
&& N = \nu  x^2  + O(x^3), \qquad x=(r-r_h)/r_h \,,\label{expNeAF}\\
&& \varphi = \varphi_h + \mu   x^n + O(x^{n+1})\,, \label{expPeAF}
 \eea
where $\mu$ and $\nu$ are dimensionless parameters, and $n$ is an
integer. Substituting this into the Eq. (\ref{eqN}) we find
 \be
\delta = \delta_h + \frac{\mu^2 n^2}{(2n-1)} x^{2n-1} + O(x^{2n})
\,,
 \ee and therefore $\e^\delta$ is finite and continuous at the
horizon. So the leading term of the l.h.s. of the Eq.
 (\ref{eqP}) is
  \be \left (N \varphi' \e^\delta r^{2} \right )' =
 \nu \mu n (n+1)\e^{\delta_h} r_h^{2}x^n  + O(x^{n+1}) \label{+6}\;.
  \ee
 This is zero at $x=0$ for any $n$,
so looking at the r. h. s. of Eq.
 (\ref{eqP}) we immediately find that  $\varphi_h=0$. Then the linear in $x$ term
at the r. h. s. of (\ref{eqP}) agrees with that at the l.h.s. and we
find:
 \be \nu n(n+1) =  {4 a^2 |QP|}/{r_h^{2}}.\lb{+8}
 \ee
 Now
consider the equation (\ref{eqD}) in the vicinity of  $r \sim r_h$.
One sees that the l.h.s. is linear in $x$, so the r.h.s. has to be
linear in $x$ too, leading to an equation for $r_h$:
 \be r_h^2=2|QP|\,, \label{+10}
  \ee
while expanding $ r^{-2} = r_h^{-2} (1-2x ) +...$ and equating the
linear in $x$ terms we obtain $\nu =1 $. Substituting this and
(\ref{+10}) into (\ref{+8}), we arrive at
 \be a^2 = \frac{n(n+1)}{2}
\,.\label{aquant}
 \ee This is the necessary condition for existence
of the AF regular extremal dilatonic dyons which coincides with
(\ref{an}). It shows, in particular, that such solutions do not
exist for $a<1$. Numerical integration show that dyons with higher
$n$ do exist as global solutions indeed \cite{Gal'tsov:2014fpa}.
This selects the class of EMD theories with discrete dilaton
couplings. Their supersymmetric origin (if any) is still enigmatic.

The set of EMD theories admitting extremal dyons exist in any
dimensions and admit generalizations with negative cosmological
constant $\Lambda$. In this case the condition on the dilaton
coupling contains one integer, one continuous parameter ($\Lambda$),
and the topological parameter $k=0,\,\pm 1$ standing for plane,
spherical and hyperbolic topologies.

\bigskip
\noindent {\bf Acknowledgments.} This work was supported by the
Russian Foundation for Fundamental Research under Project
14-02-01092.


\begin{thebibliography}{99}

%\cite{Gibbons:1982ih}
\bibitem{Gibbons:1982ih}
  G.~W.~Gibbons,
  %``Antigravitating Black Hole Solitons with Scalar Hair in N=4 Supergravity,''
  Nucl.\ Phys.\ B {\bf 207}, 337 (1982).

\bibitem{Gibbons:1987ps}
  G.~W.~Gibbons and K.~-i.~Maeda,
  %``Black Holes and Membranes in Higher Dimensional Theories with Dilaton Fields,''  Nucl.\ Phys.\ B {\bf 298}, 741 (1988).  %%CITATION = NUPHA,B298,741;%%  %877 citations counted in INSPIRE as of 18 Dec 2013
Nucl.\ Phys.\ B {\bf 298}, 741 (1988).

\bibitem{GHS}
  D.~Garfinkle, G.~T.~Horowitz and A.~Strominger,
  %``Charged black holes in string theory,''
  Phys.\ Rev.\ D {\bf 43}, 3140 (1991)
  %[Erratum-ibid.\ D {\bf 45}, 3888 (1992)].

\bibitem{GiWi}
  G.~W.~Gibbons and D.~L.~Wiltshire,
  %``Black Holes in Kaluza-Klein Theory,''
  Annals Phys.\  {\bf 167}, 201 (1986)
  [Erratum-ibid.\  {\bf 176}, 393 (1987)].

%\cite{Gibbons:1994vm}
\bibitem{Gibbons:1994vm}
  G.~W.~Gibbons, G.~T.~Horowitz and P.~K.~Townsend,
  %``Higher dimensional resolution of dilatonic black hole singularities,''
  Class.\ Quant.\ Grav.\  {\bf 12}, 297 (1995)
  [hep-th/9410073].

%\cite{Clement:1986bt}
\bibitem{Clement:1986bt}
  G.~Clement,
  %``Rotating {Kaluza-Klein} Monopoles and Dyons,''
  Phys.\ Lett.\ A {\bf 118}, 11 (1986).

%\cite{Rasheed:1995zv}
\bibitem{Rasheed:1995zv}
  D.~Rasheed,
  %``The Rotating dyonic black holes of Kaluza-Klein theory,''
  Nucl.\ Phys.\ B {\bf 454}, 379 (1995)
  [hep-th/9505038].
  %%CITATION = HEP-TH/9505038;%%

\bibitem{Galtsov:1995mb}
 D.~V.~Galtsov, A.~A.~Garcia and O.~V.~Kechkin,
  %``Symmetries of the  stationary Einstein-Maxwell dilaton theory,''
  Class.\ Quant.\ Grav.\  {\bf 12}, 2887 (1995)
  [arXiv:hep-th/9504155].
  %%CITATION = CQGRD,12,2887;%%

%\cite{Poletti:1995yq}
\bibitem{Poletti:1995yq}
  S.~J.~Poletti, J.~Twamley and D.~L.~Wiltshire,
  %``Dyonic Dilaton black holes,''
  Class.\ Quant.\ Grav.\  {\bf 12}, 1753 (1995)
  [Erratum-ibid.\  {\bf 12}, 2355 (1995)]
  [hep-th/9502054].

%\cite{Chen:2008hk}
\bibitem{Chen:2008hk}
  C.~-M.~Chen, D.~V.~Gal'tsov and D.~G.~Orlov,
  %
  %``Extremal dyonic black holes in D=4 Gauss-Bonnet gravity,''
  Phys.\ Rev.\ D {\bf 78}, 104013 (2008)
  [arXiv:0809.1720 [hep-th]].
%\cite{Gibbons:1993xt}
\bibitem{Gibbons:1993xt}
  G.~W.~Gibbons, D.~Kastor, L.~A.~J.~London, P.~K.~Townsend and J.~H.~Traschen,
  %``Supersymmetric selfgravitating solitons,''
  Nucl.\ Phys.\ B {\bf 416}, 850 (1994)
  [hep-th/9310118].
\bibitem{Nozawa:2010rf}
  M.~Nozawa,
  %``On the Bogomol'nyi bound in Einstein-Maxwell-dilaton gravity,''
  Class.\ Quant.\ Grav.\  {\bf 28}, 175013 (2011)
  [arXiv:1011.0261 [hep-th]].

%\cite{Wiltshire:1994de}
\bibitem{Wiltshire:1994de}
  D.~L.~Wiltshire,
  %``Dilaton black holes with a cosmological term,''
  J.\ Austral.\ Math.\ Soc.\ B {\bf 41}, 198 (1999)
  [gr-qc/9502038].

%\cite{Poletti:1994ww}
\bibitem{Poletti:1994ww}
  S.~J.~Poletti, J.~Twamley and D.~L.~Wiltshire,
  %``Charged dilaton black holes with a cosmological constant,''
  Phys.\ Rev.\ D {\bf 51}, 5720 (1995)
  [hep-th/9412076].
\bibitem{HoHo}
J.H. Horne and G.T. Horowitz,
%``Black holes coupled to a massive dilaton'',
Nuclear Physics {\bf B399} (1993) 169--196 [hep-th/9210012].

%\cite{Gal'tsov:2014fpa}
\bibitem{Gal'tsov:2014fpa}
  D.~Gal'tsov, M.~Khramtsov and D.~Orlov,
  %``"Triangular" extremal dilatonic dyons,''
  Phys.\ Lett.\ B {\bf 743}, 87 (2015)
  [arXiv:1412.7709 [hep-th]].


  %%CITATION = GR-QC/9502038;%%
  %9 citations counted in INSPIRE as of 18 Dec 2013

\end{thebibliography}
\end{document}